# Enhancing Password Security Through a High-Accuracy Scoring Framework Using Random Forests

Muhammed El Mustaqeem Mazelan
*Universiti Kuala Lumpur*
muhammed.mazelan03@s.unikl.edu.my

*Noor Hazlina Abdul Mutalib
*Universiti Kuala Lumpur*
noor.hazlina@unikl.edu.my
*Corresponding author

*Nouar AlDahoul
*New York University*
*Abu Dhabi*
naa9497@nyu.edu
*Corresponding author

Password security plays a crucial role in cybersecurity, yet traditional password strength meters, which rely on static rules like character-type requirements, often fail. Such methods are easily bypassed by common password patterns (e.g., 'P@ssw0rd1!'), giving users a false sense of security. To address this, we implement and evaluate a password strength scoring system by comparing four machine learning models: Random Forest (RF), Support Vector Machine (SVM), a Convolutional Neural Network (CNN), and Logistic Regression with a dataset of over 660,000 real-world passwords. Our primary contribution is a novel hybrid feature engineering approach that captures nuanced vulnerabilities missed by standard metrics. We introduce features like leetspeak-normalized Shannon entropy to assess true randomness, pattern detection for keyboard walks and sequences, and character-level TF-IDF n-grams to identify frequently reused substrings from breached password datasets. our RF model achieved superior performance, achieving 99.12% accuracy on a held-out test set. Crucially, the interpretability of the Random Forest model allows for feature importance analysis, providing a clear pathway to developing security tools that offer specific, actionable feedback to users. This study bridges the gap between predictive accuracy and practical usability, resulting in a high-performance scoring system that not only reduces password-based vulnerabilities but also empowers users to make more informed security decisions.

## 1. INTRODUCTION

Passwords remain a cornerstone of online security, serving as the primary means of authentication for countless systems and applications. However, this reliance is a critical vulnerability; according to a report by Google Cloud, a staggering 86% of breaches involve stolen credentials, posing a significant threat to both user data and system security.[1] Many users choose weak, easily guessable passwords, which pose a serious threat to both user data and system security. [2] reported that users often create passwords through predictable patterns such as dictionary words, keyboard sequences, or personal information are highly vulnerable to modern guessing models. Attackers frequently exploit this vulnerability in large-scale attacks, compromising user privacy and enabling financial fraud. Most traditional password strength scoring tools rely on static rules, such as requiring a mix of lowercase, uppercase, digits, and special characters (LUDS), which fail to adapt to evolving attack patterns. As highlighted in [3], these rule-based systems often overlook the structural and semantic patterns in user-chosen passwords, such as dictionary words, character position biases, and common mangling rules (e.g., replacing 'a' with '@' or appending digits). The paper argues that such simplistic metrics do not adequately reflect the actual resistance of a password against modern guessing attacks, particularly probabilistic methods like Markov models or context-free grammars.

To overcome the limitations of static and rule-based checkers, we introduce a password scoring system built on a machine learning framework. Our approach does not disregard fundamental metrics like password length and character diversity. It treats them as part of a larger and interconnected feature space. The central innovation of our system is its use of a non-linear model to analyze the complex interplay between these basic metrics and more sophisticated features, such as predictable keyboard patterns and dictionary word inclusion. A traditional checker might simply award points for length and special characters independently. In contrast, our model can learn that a long password consisting of a dictionary word followed by a simple numeric sequence is fundamentally weaker than a shorter, more random password with a distinction requiring contextual understanding of features, rather than isolated analysis. The feature set engineered for this purpose includes password length, entropy, character diversity, pattern detection, and dictionary matching, enabling a holistic analysis that adapts to evolving attack patterns and provides a more accurate measure of password strength.

The remainder of this paper is organized as follows. Section II reviews the related work on evolution of password strength assessment methodologies. Section III details our proposed methodology, including the dataset, feature engineering, and model architectures. Section IV presents and discusses the experimental results, and Section V concludes the paper with a summary of our findings and directions for future work.

## 2. Related work

This section reviews the evolution of password strength assessment methodologies to contextualize our work. We examine three primary paradigms: traditional heuristic-based approaches, information-theoretic models, and modern machine learning techniques, critically evaluating their limitations to establish the research gap addressed by our work.

Authentication systems verify the identity of users or devices and serve as the first line of defense against unauthorized access to secure systems. Among various methods, password-based authentication remains one of the most common authentication methods used in computer systems because it is easy to implement, cost-effective, and familiar to users. [4] Password security refers to the procedures and policies that protect passwords from misuse, unauthorized access, or exposure to online threats. The persistent issues associated with password-based authentication have driven ongoing innovation in password strength assessment approaches. These methods have evolved from simple rule-based systems to more advanced probabilistic models such as Markov-based password strength meters, and even to recent developments using machine learning techniques. [5]

### 2.1. Traditional and Rule Based Password Strength Meters (PSM)

Early versions of Password Strength Meters (PSMs) relied heavily on heuristic rules, such as requiring passwords to include characters from different groups (e.g., uppercase, lowercase, digits, special characters). Some rule-based methods also compare passwords against large sets of common terms to assess guess ability. The paper notes that Commercial PSMs, such Google Password Meter (GPM), Microsoft Password Checker (MPC), and The Password Meter, often use lexical analysis, which looks at English words and phrases [6].

However, traditional and rule-based password strength meters (PSMs) have significant limitations. Because they rely on heuristic approaches, they are often inflexible and tailored to defend against outdated attacks like simple brute-force or dictionary attacks using static lists of common passwords, making them less effective against evolving threats. [7] Consequently, their effectiveness is limited, and users may choose passwords that receive high scores from these meters but still provide little real protection against modern password cracking techniques [8]. Additionally, their focus on lexical analysis of English words makes them impractical for users in non-English-speaking regions, [9] and many commercial password checkers require subscription fees, reducing their accessibility.

Overreliance on traditional password strength significantly threatens system security. Evolving user habits and attack tactics have rendered traditional evaluation criteria such as minimum length, minimum character set, number of characters, and uppercase have become ineffective. [10] Previous studies indicate that, although users often comply with password requirements such as including digits, uppercase letters, or special characters by making simple modifications (like adding numbers or substituting letters), their passwords frequently remain weak and susceptible to modern attacks, including brute-force and guessing attacks. This leads to passwords that may appear strong but are still easily compromised by systematic attacks and advanced tools. [11] As attackers exploit these predictable patterns, systems become increasingly vulnerable, and traditional evaluation methods lose effectiveness. Existing password strength scoring approaches often fail to provide accuracy in terms of feedback in real-time applications, [12] especially when they do not consider relationship between user passwords and sensitive personal information, which can pose a serious security risk. Moreover, some methods may offer more realistic feedback but are not suitable for real time use due to its excessive computation time. [13] In this study, we evaluate the strengths and limitations of Random Forest (RF), Convolutional Neural Networks (CNN), Support Vector Machines (SVM), and Logistic Regression (LR) as machine learning models.

### 2.2. Entropy and Information-Theoretic Approaches

Entropy-based methods provide a quantitative approach to evaluating password strength. These methods assess factors such as password length, complexity, and patterns to determine how strong a password is. In this paper, machine learning algorithms analyze various factors including length, complexity, entropy, and patterns to estimate password strength more accurately than traditional heuristic-based methods [14]. The approach aims to offer a more refined and effective assessment of password robustness compared to conventional metrics.

Even though many major digital platforms use National Institute of Standards and Technology (NIST) entropy, it has been thoroughly tested to see how well it works in practice. However, research has shown that this entropy-based technique only delivers an inadequate approximation of password strength when evaluated against actual password guessing attacks. [15] As a result, "guessability," which measures the time or computational effort required for a cryptanalysis algorithm to recover a password, is now seen as a more convincing assessment of password strength. This suggests that theoretical measures of randomness may not accurately reflect a password's real-world resistance to attacks. [16]

## 2.3. Machine learning and deep learning

Machine learning offers a potential solution by enabling systems to analyze large databases, which allows systems to examine massive databases of actual passwords, identify subtle trends, and give more accurate evaluations of password strength. [17] Researchers have found promising results using password classification algorithms such as Random Forest, Support Vector Machines (SVMs), and boosting techniques when it comes to identifying weak, medium, and strong passwords [18]. Additionally, Deep learning methods, such as Long Short-Term Memory (LSTM) networks and Generative Adversarial Networks (GANs), have shown advantages in extracting password patterns and generating resilient passwords, outperforming traditional rule-based approaches in both password guessing and password strength evaluation [19].

Recent research has introduced innovative frameworks that combine machine learning with advanced feature extraction methods. For example, the use of Term Frequency-Inverse Document Frequency (TF-IDF) for feature extraction, combined with Random Forest classification, has achieved accuracy rates as high as 95% in password strength evaluation tasks [20]. Beyond basic requirements such as character length or type, these models can identify password vulnerabilities including susceptibility to dictionary attacks or patterns that resemble human behaviour.

TABLE 1. COMPARE HYBRID FEATURE SET AND THE RESULTING ACCURACY TO A FEW OTHER KEY PAPERS THAT HAVE ALSO USED MACHINE LEARNING FOR PASSWORD STRENGTH EVALUATION.

| Author | Methods | Strengths | Limitations |
|---|---|---|---|
| Abu Asaduzzaman, Declan D'souza, Raihan Uddin, Yoel Woldeyes (2024) | Uses Term Frequency-Inverse Document Frequency (TF-IDF) to analyze character frequency, training a logistic regression model on a leaked password list. | Develops a practical machine learning model that provides a data-driven way to recommend stronger passwords. | The study does not explore more advanced models or ensemble techniques for potentially higher accuracy. |
| Shreya, K., Divakarla, L. P., Kumar, K. J., & Vishwas, H. N. (2024). | The study employs a supervised machine learning approach to classify passwords into 3 categories based on features like length, character diversity, entropy, and pattern detection, with TF-IDF used for feature extraction and LIME for explainability. | Achieves high accuracy (95% with Random Forest), uses explainable AI (LIME) to interpret model predictions, incorporates multiple robust ensemble models, and focuses on real-time password strength assessment to guide users toward stronger passwords. | Relies on a pre-existing dataset from Kaggle which may not reflect current password trends, does not address password reuse or behavioral patterns, and lacks integration into a live system for real-world validation. |
| Binh Le Thanh Thai And Hidema Tanakaa (2024) | Proposes new hybrid Markov models that consider both character position and password length to calculate password probability. | Offers a detailed comparative analysis of different Markov models, showing improved accuracy with the proposed hybrid approach. | The focus is on Markov models, which might not capture all types of password weaknesses like those found in dictionary attacks. |
| R. Rathi, P. Visvanathan, R. Kanchana and R. Anand (2020) | Performs a comparative analysis of several soft computing techniques, including Back Propagation Neural Network (BPN), Logistic Regression (LR), and Convolutional Neural Network (CNN). | Provides a comprehensive overview and comparison of multiple soft computing methods for the task of password classification. | The paper's primary focus is on comparison rather than proposing a single, novel, or highly optimized technique. |
| Rengkung, B. J., Royan, N., & Roestam, R. (2024). | Combines K-Means clustering for feature extraction with an entropy-based classification method to categorize passwords. | Achieves high accuracy (over 99%) by using a unique combination of supervised and unsupervised learning techniques. | The dataset used for training and testing is relatively small (5,332 passwords). |
| Nosenko, A., Cheng, Y., & | Employs a Recurrent Neural | Introduces a deep learning approach | Requires substantial computational |

| Chen, H. (2022) | Network (RNN) with Long Short-Term Memory (LSTM) units, specifically using the GPT-2 Small model (124M parameters), fine-tuned on password datasets to predict user password modifications and guess both passwords and passphrases | capable of learning complex password modification patterns, effectively models passphrases using natural language corpora, outperforms traditional rule-based and Markov models in guessing efficiency, and provides insights into user password change behavior. | resources for training deep models, relies on large datasets for effective performance which may raise privacy concerns, results may be biased toward English-language patterns, and the model's practical deployment for real-time password strength assessment is not demonstrated. |
|---|---|---|---|
| R. Divya, Gaganashree, S. B. Devamane, V. Dharshini and S. Deepika 2023 | Evaluates the performance of several supervised machine learning for password strength classification into 3 categories; preprocesses the data by removing null values, splits it into training and testing sets, and assesses model performance. | Compares multiple classifiers systematically, uses a clear and reproducible evaluation framework with standard metrics, identifies Random Forest as the top-performing model (98.7% accuracy), and provides a straightforward workflow for password strength assessment that can be integrated into security systems. | Relies on a dataset generated by a third-party password strength tool which may introduce labeling bias, does not use real-world leaked password data, lacks advanced feature engineering or NLP-based techniques, and does not incorporate explainability or real-time user feedback mechanisms. |

In this study, we designed a password strength scoring system based on machine learning. Our proposed system aims to overcome the shortcomings of conventional methods and offer a complete, adaptable answer to the problem of assessing and improving password security by drawing on the findings of previous studies and incorporating more sophisticated algorithms.

Feature engineering is also a key area of research. For instance, some frameworks have combined TF-IDF feature extraction with Random Forest classifiers to achieve high accuracy in password strength verification.

Our goal is to identify the most practical real-time password scoring model by comparing their accuracy using criteria such as recall, precision, and F1-score. Currently, users often do not adopt secure password habits and frequently create passwords that are easily guessable, which presents significant security risks. This highlights the need for improved systems that not only assess password strength but also educate users on creating stronger passwords and reducing predictable patterns in their behavior [21]. For example, a system should be able to suggest passphrases that comply with NIST standards, making them more secure. [22] The objectives of this study are:

1. To study existing password strength scoring system and their limitations.
2. To develop a machine learning-based password strength scoring system that more accurately predicts a password's resistance to cracking attempts.
3. To test the performance of the system compared to traditional password strength scoring.

This study focuses on developing a machine learning-based password strength evaluation system and testing its effectiveness using the following key features:

1. Target Users: We designed this system for individuals and organizations aiming to improve password security, especially those vulnerable to cyberattacks due to weak or reused passwords.
2. Functions and Features: Our system classifies passwords as *weak, medium, or strong*, provides real-time feedback, and offers actionable recommendations to improve password strength. We trained the underlying machine learning models on diverse datasets.
3. Implementation Platform: We've built our system as a Python library, and we're currently developing and testing it using Python and Google Colab. To make it highly flexible and easy to use, is also being developed as an API. By offering our password strength scoring system as an API, it can be effortlessly integrated into all sorts of web applications and platforms.

To address this research gap, our work introduces a novel hybrid feature engineering approach that integrates three distinct paradigms of analysis. Unlike prior models that rely primarily on surface-level metrics, our feature set is designed for deeper vulnerability detection. Specifically, our contributions are:

1. We directly address the limitation of standard entropy by first normalizing common leetspeak substitutions, providing a more accurate measure of a password's true randomness.
2. Moving beyond simple dictionary matching, we employ TF-IDF on character n-grams. This data-driven technique identifies frequently reused substrings and patterns learned from a large corpus of breached passwords, capturing contextual vulnerabilities that are otherwise missed.
3. We implement and rigorously compare four different ML models on a massive, real-world dataset to identify the best balance of accuracy and efficiency.

4. We introduce a novel hybrid feature set that combines the strengths of all three paradigms: basic metrics (length), information theory (leetspeak-normalized entropy), and data-driven pattern analysis (TF-IDF n-grams).

## 2.4. Common Issues With Password Security

*Structural Predictability from Rule-Based Complexity Requirements*:
Many password policies require basic composition rules but fail to address deeper structural vulnerabilities. Users commonly meet these requirements using predictable substitutions. For instance, a password that use numbers for some of the letters in a word or name such as 'P@ssw0rd123!' are still highly guessable, with 54% likely to be guessed. [23] This password would likely be rated as 'strong' by many traditional checkers because it satisfies requirements for length, uppercase letters, digits, and special characters. However, this password is built on a common dictionary word with predictable character substitutions and a sequential numeric suffix, making it highly vulnerable to modern hybrid dictionary and rule-based attacks. This study aims to develop a model capable of identifying these nuanced vulnerabilities that evade simple static rules. These practices create passwords that meet policy requirements but remain highly susceptible to modern, pattern-based attacks. The threat has been amplified by modern machine learning models, such as PassGAN, which can learn and replicate these common user-chosen password structures. This underscores the need for scoring systems that can recognize these nuanced patterns [24]. Our inclusion of dedicated pattern detection and character-level n-gram analysis features is a direct response to this threat, enabling our model to identify the subtle structural regularities that make a password predictable and vulnerable to advanced guessing tools.

*Password Reuse*: Users continue to reuse passwords across multiple accounts, creating critical vulnerabilities in authentication systems. As noted by (Asaduzzaman et al., 2024) [25] password reuse is a huge security risk because one service's data breach may create a domino effect leading to larger breaches. Similarly, (Xiao & Zeng, 2022) [26] highlight that password reuse across accounts is common, and a leak from one platform can jeopardize the security of others. Furthermore, large-scale analyses of leaked datasets confirm that password reuse and predictable patterns are persistent issues that amplify the impact of single-platform breaches. Analysis of over 1 billion leaked passwords revealed that 84.31% of passwords contain lowercase letters, 69.64% contain numbers, and 25.87% consist of lowercase letters only, highlighting predictable patterns that increase susceptibility to cross-service password exploitation [27]. The LinkedIn data breach demonstrated that even passwords rated as "strong" by commercial checkers are vulnerable, with up to 15.40% being cracked using advanced algorithms when attackers leverage password strength checkers to select training data [28]. Although some of users recognize the risks, they continue this dangerous behavior due to convenience and memorability concerns. Even though users are aware of the importance of password security, they often choose simple and easy-to-remember passwords, prioritizing ease of recall over security and thus ignoring password strength [29]. Prior large-scale studies also have found that users tend to generate easy-to-guess passwords in certain structures and often use repeated passwords. This pattern of password reuse is well documented and contributes to the vulnerability of password-based authentication systems. [30]

*Lack Of Awareness*: Many users lack awareness of fundamental password security principles, often prioritizing memorability over robustness despite widespread educational efforts. This gap in understanding leads to the selection of passwords that, while seemingly complex, are highly predictable. Users often create passwords using common patterns, making their credentials vulnerable to attack, as many records contain weak passwords or well-known password patterns, which automated password cracking algorithms can exploit both offline and online [31]. This issue is further exacerbated by static password policies, which do not adapt to evolving attack strategies and fail to encourage users to adopt more secure behaviors [32]. As a result, traditional rule-based password checkers that focus mainly on requirements like length and character variety, often fail to detect weak passwords because these checkers rely on simple rules and do not account for the true complexity or unpredictability of user-chosen passwords, highlighting the limitations of static, rule-based approaches. [33]

## 3. METHODOLOGY

In this study, we apply a supervised machine learning approach to classify password strength into three distinct categories: weak, medium, and strong. Our methodology encompasses a comprehensive, multi-stage framework designed for robust and accurate password scoring. This framework covers the selection and preprocessing of a large-scale, real-world dataset; a novel hybrid feature engineering process to capture both structural and semantic vulnerabilities; the implementation and training of four machine learning models (Random Forest, SVM, CNN, and Logistic Regression); and a rigorous evaluation framework to compare their performance based on key security-oriented metrics. We preprocessed the data using Synthetic Minority Oversampling Technique (SMOTE) to address class imbalance and features were normalized or vectorized as required. Each model was trained and validated using a stratified split, and evaluated with accuracy, precision, recall, and F1-score to ensure robustness and generalization.

## 3.1. Dataset Used

In this study, we analyzed passwords from the 000webhost leak, categorized by the PARS tool from Georgia Tech, using commercial metrics from Twitter, Microsoft, and Battle.net, along with strength labels (0: weak, 1: medium, 2: strong). [34] While this dataset provides a valuable large-scale resource for training, this represents a potential limitation, as our models are trained to replicate the latent rules of this original labeling scheme. Future work could involve re-labeling the dataset using modern cracking tools to establish a ground truth based purely on guessability.

The dataset contains 669,639 passwords and includes raw passwords and pre-annotated features. A crucial characteristic of this dataset is its significant class imbalance. An analysis of the ground-truth labels revealed the following distribution:

- Weak: 156,000 passwords (~23.3%)
- Medium: 432,000 passwords (~64.5%)
- Strong: 80,000 passwords (~12.2%)

We validated the dataset and its quality through the following preprocessing steps:

a. Removed duplicates and null values.

b. Standardized passwords: This step involved two key actions. First, all alphabetic characters were converted to lowercase to ensure case-insensitivity for feature analysis like dictionary matching. Second, special characters were explicitly retained and consistently encoded (UTF-8). This ensured that special characters were available for downstream feature extraction (such as character counts and entropy calculation) rather than being removed or inconsistently processed.

c. Addressed class imbalance using SMOTE: To mitigate the bias from the skewed class distribution, we applied the Synthetic Minority Oversampling Technique (SMOTE). This technique was applied only to the training data to prevent data leakage, balancing the classes by creating synthetic examples of the minority classes. This ensures that the model gives equal importance to weak, medium, and strong passwords during the training phase.

## 3.2. Feature Engineering

To capture the complexity and vulnerabilities of passwords, we designed a hybrid set of features combining basic string metrics, semantic patterns, and advanced entropy-based analysis. These features reflect both structural weaknesses, such as short length and predictable sequences, and semantic risks such as dictionary words and common leetspeak substitutions. Below is a formal breakdown of the key engineered components.

1. *Basic String Metrics*
   a. Length: The total number of characters in each password. Shorter passwords (e.g., <8 characters) are inherently weaker due to a smaller search space for brute-force attacks.
   b. Character Counts: We counted the occurrences of characters from four distinct classes:
      - Digits (e.g., 1, 2).
      - Uppercase letters (e.g., A, B).
      - Lowercase letters (e.g., a, b).
      - Special characters (e.g., !, @, $)

These counts help identify compliance with traditional LUDS (Lowercase, Uppercase, Digits, Symbols) policies but are primarily used as inputs for the more sophisticated features below.

2. *Structural and complexity metrics*

These metrics were designed to quantify the true complexity of a password beyond simple character counts.

A composite metric to quantify the diversity of character types used. The score is calculated as the total number of unique character classes present in the password.

Let I(class) be an indicator function that returns 1 if at least one character from that class is present, and 0 otherwise. The classes are defined as lowercase, uppercase, digits, and special characters. The score is calculated using the following formula:

$$Score = I(lowercase) + I(uppercase) + I(digits) + I(special) \quad (1)$$

The score ranges from 1 (e.g., 'password') to 4 (e.g., 'P@ssw0rd!').

3. *Normalized Entropy*

To assess the true, unpredictable information content of a password, we calculated its Shannon entropy after a leetspeak normalization step. This process is crucial to avoid overestimating the strength of passwords that use common, predictable substitutions.

Step 1: Leetspeak Normalization. Prior to entropy calculation, each password string underwent a normalization process where common leetspeak characters were converted back to their base alphabetical equivalents. The substitutions selected for this study represent the most frequently observed and predictable character replacements seen in real-world passwords. This list was curated based on extensive analysis in prior cybersecurity literature and password breach datasets, targeting the substitutions that are most intuitive for users due to their visual similarity to the letters they replace. This targeted approach ensures that the most common methods for artificially inflating password complexity are neutralized before entropy calculation. The specific substitutions applied were:

@, 4 → a
$, 5 → s
1, ! → l
0 → o
3 → e
7 → t

Step 2: Shannon Entropy Calculation. On the normalized string, the Shannon entropy H(X) was calculated using the formula:

$$H = -\Sigma(p(i) * \log_2(p(i))), \qquad (2)$$

where p(i) is the normalized password string, i represents a unique character in the string, and p(i) is the probability of that character's occurrence, calculated as the character's frequency divided by the total length of the string. This normalization ensures that a password like 'P@ssw0rd' is correctly analyzed based on the lower entropy of its root, 'password'.

4. *Dynamic Character Set Size*

This metric adjusts the potential character pool size based on the types of characters actually present in the password, providing a more realistic basis for theoretical strength. The total character set size N was calculated as the sum of the pool sizes for each character class present in the password. The defined pools were:
Lowercase letters (a-z): Pool size = 26
Uppercase letters (A-Z): Pool size = 26
Digits (0-9): Pool size = 10
Special characters (~!@#$%^&*()_-+={[}]|\:;"'<,>.?/):
Pool size = 32

For example, a password like 'Password123' would have a dynamic character set size N of 26 + 26 + 10 = 62. A password like 'P@ssw0rd!' (containing all four classes) would have N = 26 + 26 + 10 + 32 = 94.

5. *Pattern Detection*
- Sequential Patterns: Identified keyboard walks (e.g., 1234, qwerty) or alphabetical/numerical sequences (e.g., ABC, 789).
- Repeated Characters: Flagged passwords that contained three or more identical characters (e.g., aaaa, 1111) or repeated digits (e.g., 123123).

6. *Semantic Analysis*
- Dictionary Word Matches

To identify semantically weak passwords, we checked each password for substrings that matched a curated list of common dictionary words and breached passwords. For this study, a curated list of the top 197 most breached passwords from 2023/2024 was used for semantic analysis. We acknowledge that this list is not exhaustive and represents a limitation of the current implementation. While effective for identifying the most common and trivially guessable passwords, future work should integrate larger dictionaries (e.g., the multi-million entry RockYou corpus) to enhance the detection of less common but still compromised passwords.

7. *Hybrid Feature Integration*
- TF-IDF N-Gram Vectors: Combined the previous features with character-level n-grams (1–2 characters) to capture spatial dependencies (e.g., 12, qwerty). This added contextual awareness of frequent substrings in breached datasets.
- Feature Scaling: Standardized numerical features (e.g., length, entropy) and sparse n-gram vectors for model compatibility.

### 3.3. Model Architecture and Implementation

We developed a machine learning-based Password Strength Scoring System to address critical issues such as password reuse and user unawareness. Unlike static rule-based checkers, the system leverages a hybrid ML framework comprising Random Forest (RF), Support Vector Machine (SVM), and Convolutional Neural Network (CNN), with logistic regression (LR) as a baseline model. We configured the RF model with 100 decision trees to capture non-linear feature interactions using Gini impurity, with entropy and length emerging as dominant predictors. We trained the SVM model with an RBF kernel, and it demonstrated strong generalization. We built the CNN model using a 1D architecture with character-level embeddings, convolutional layers, and max pooling, optimized for detecting local patterns but less effective on short passwords. The logistic regression (LR) model offered interpretability but showed limited accuracy.

All models were trained on a hybrid feature set comprising password string metrics, entropy calculations, pattern detection, and dictionary-based analysis. The entire experimental pipeline, from data preprocessing to model evaluation, was implemented in Python 3.8. Data manipulation and feature engineering were conducted primarily using the Pandas and NumPy libraries. For the machine learning models, we utilized the Scikit-learn library for the Random Forest, SVM, and Logistic Regression implementations, as well as for feature scaling, data splitting, and performance metric calculations. The SMOTE implementation was sourced from the imbalanced-learn library. The Convolutional Neural Network (CNN) was built and trained using the TensorFlow framework with its Keras API.

The dataset, annotated with three strength classes (weak, medium, strong), was stratified and split into training (70%), validation (10%), and test (20%) sets. SMOTE was applied to the training data to mitigate class imbalance. Hyperparameters for each model were optimized using grid search with cross-validation. For the

CNN, early stopping with validation loss monitoring was implemented to prevent overfitting. All features were scaled or vectorized appropriately, ensuring compatibility across all models.

All models were evaluated using accuracy, weighted F1-score, and confusion matrices to address class imbalance (weak: 156k, medium: 432k, strong: 80k). The Random Forest achieved the highest test accuracy (99.12%) and F1-score (0.9911), excelling in distinguishing weak passwords via entropy and length thresholds. The SVM closely followed (99.02% accuracy), while the CNN lagged slightly (93.95%) due to challenges with short passwords (<8 characters). Logistic Regression served as a baseline, attaining 90.22% accuracy but struggling with recall for weak passwords (62%).

Hyperparameter tuning and cross-validation ensured robustness, grid search optimized tree depth and kernel parameters, while SMOTE oversampling balanced class distributions. The final Random Forest model was selected for deployment due to its balance of accuracy (99.12%), interpretability, and computational efficiency, making it suitable for real-time feedback in applications like password managers or authentication APIs.

### 4. RESULTS AND DISCUSSION

#### 4.1. Comparative Model Performance

Our experimental results show that the Random Forest (RF) model achieved the highest test accuracy (99.12%) and weighted F1-score (0.9911), outperforming both the SVM (99.02% accuracy, 0.9902 F1) and CNN (93.95% accuracy, 0.9395 F1). The CNN's reduced overall performance can be attributed to its challenges with short passwords (<8 characters), which constitute a large portion of the 'weak' password class. A granular analysis revealed that the CNN achieved a recall of only 71% for 'weak' passwords, meaning it failed to identify nearly a third of the most vulnerable entries. In stark contrast, the Random Forest model maintained a recall of 99.2% for the same class, highlighting its superior ability to leverage the engineered features even when the spatial patterns required by convolutional layers are minimal.

#### 4.2. Feature Correlation Analysis

To pinpoint the most crucial features for classifying password strength, we employed SelectKBest with the ANOVA F-value scoring function, which assesses how effectively each feature differentiates between strength classes. Our results revealed that character variety scores and entropy were the most influential predictors, highlighting their essential role in distinguishing weak from strong passwords due to their inherent resistance to brute-force and dictionary attacks. Pattern detection emerged as the fifth most significant feature, emphasizing that even seemingly complex passwords can be vulnerable if they contain predictable structural patterns like "123456" or "password123". In contrast, basic metrics received the lowest scores, indicating that simple character counts alone are insufficient for accurate strength assessment, a finding consistent with criticisms of traditional LUDS (Lowercase, Uppercase, Digits, Symbols) password policies. These findings highlight the value of using a hybrid feature engineering approach, combining classical metrics like entropy with semantic and structural checks such as pattern detection, to identify nuanced vulnerabilities that rule-based systems often miss.

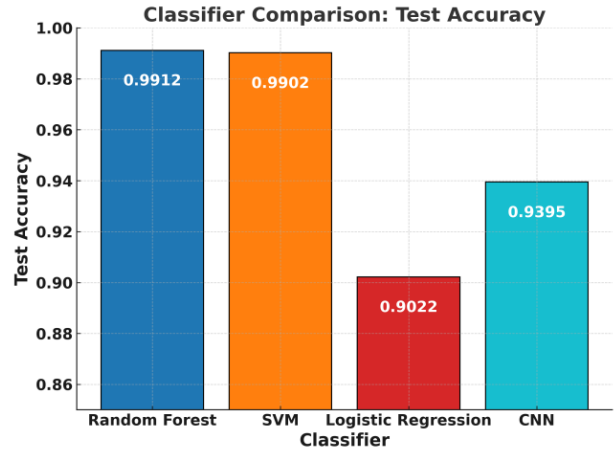

FIG. 1. CLASSIFIER COMPARISON ON TEST ACCURACY

Table 2 summarizes the comparative performance metrics of the models.

TABLE 2. Comparative Model Performance Metrics

| Model | Accuracy | Precision | Recall | F1-Score |
|---|---|---|---|---|
| RF | 0.9912 | 0.99 | 0.99 | 0.99 |
| SVM | 0.9902 | 0.99 | 0.99 | 0.99 |
| LR | 0.9022 | 0.91 | 0.90 | 0.90 |
| CNN | 0.9395 | 0.94 | 0.94 | 0.94 |

#### 4.3. Confusion Matrix

The confusion matrix in Fig. 3 provides a detailed breakdown of the Random Forest model's performance, confirming its high accuracy across all categories. The most important insight for a security tool, however, is its performance on 'weak' passwords. Of the 31,349 true weak passwords in the test set, only 47 (a mere 0.15%) were misclassified as 'strong.' This exceptionally low false negative rate is the model's most critical feature for a security application, as it directly minimizes the primary risk: providing a user with a dangerous false sense of security. This result strongly validates the model's reliability and suitability for real-world deployment where user safety is paramount.

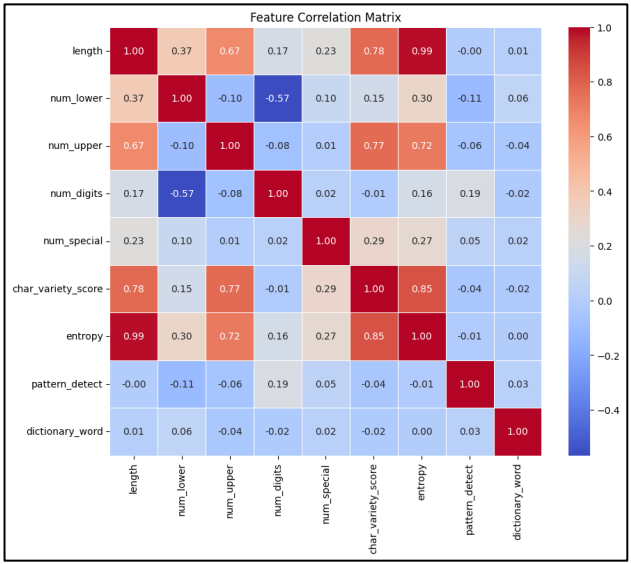

FIG. 2. FEATURE CORRELATION MATRIX

It is crucial to interpret our model's performance beyond simple classification accuracy and connect it to the practical cybersecurity concept of 'guessability.' While theoretical entropy provides a measure of randomness, guessability quantifies the real-world computational effort required for an attacker to compromise a password. In this context, our Random Forest model's high precision and recall for the 'weak' password class can be viewed as a highly effective proxy for identifying passwords with low guessability.

This assertion is strongly supported by our feature importance analysis (Figure 4). The model identified normalized entropy, character variety score, and length as the most influential predictors of password strength. These are precisely the features that most directly contribute to expanding an attacker's search space and, consequently, increasing the time and resources required to guess the password. The model's ability to assign the highest importance to these specific features demonstrates that it has not merely learned to fit the training labels but has successfully identified the fundamental statistical properties that make a password vulnerable to brute-force and dictionary attacks. Therefore, the high performance of our model is significant because it represents a practical, data-driven method for estimating a password's resistance to real-world guessing attacks.

### 4.4. Characterizing the 'Medium' Password Category

While the performance on 'weak' and 'strong' passwords defines the security-critical boundaries, an analysis of the 'medium' category (labeled as 'Good' in the confusion matrix) provides deeper insight into the model's nuanced decision-making process. This category typically represents passwords that satisfy basic complexity requirements but still contain underlying, predictable flaws. Based on our feature importance analysis, the model identifies 'medium' passwords as those exhibiting a mixture of strong and weak characteristics. For example:

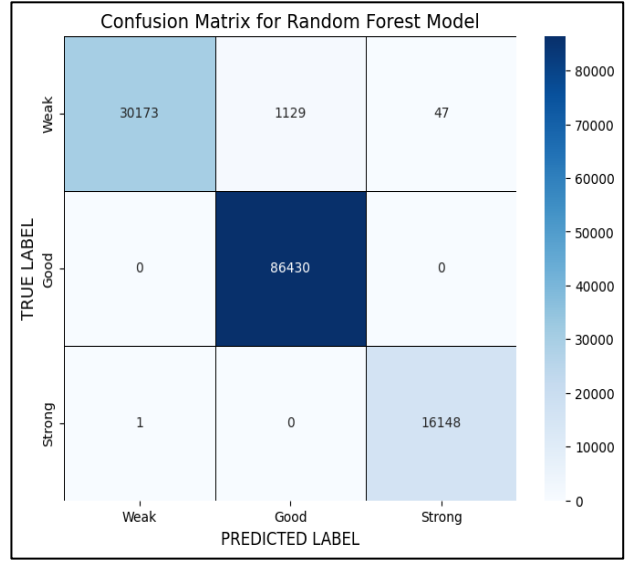

FIG. 3. CONFUSION MATRIX FOR RANDOM FOREST CLASSIFIER

a. *High Complexity, Semantic Weakness:*
A password like 'P@ssword123!' has a good length and high character variety (score of 4), but its root is a common dictionary word with predictable substitutions. Our model correctly identifies this semantic weakness, classifying it as 'medium' rather than 'strong'.
b. *Moderate Length, Low Entropy:*
 A password like 'boatboatboat1' may be sufficiently long but has low entropy due to character repetition, preventing it from being classified as 'strong'.
c. *Apparent Randomness with Hidden Patterns:*
A password may appear random but contain a keyboard walk (e.g., 'asdfg!@#$'). The pattern detection feature would flag this, likely placing it in the 'medium' category.

An examination of the misclassifications involving this category is particularly revealing, based on the confusion matrix in Fig. 3:
a. *Weak Passwords Misclassified as Medium (1,129 cases):*
This is the most common error involving the 'medium' category. These are typically "borderline" weak passwords that meet a minimum threshold for one feature, pushing them over the decision boundary. For instance, a password like 'password!' might be classified as medium because the inclusion of a single special character increases its character variety score, even though its semantic core is extremely weak. This highlights the model's sensitivity to the feature set.

*b. Strong Passwords Misclassified as Medium (1 case):*
This error is exceptionally rare, demonstrating the model's robust ability to recognize truly strong passwords. The single misclassification likely represents a password that sits precisely on the model's decision boundary, meeting most but not all of the implicit criteria for the 'strong' class.

*c. Medium Passwords Misclassified (1,129 as Weak, 47 as Strong):*
The majority of errors demote 'medium' passwords to 'weak'. This indicates that when the model is uncertain, its decision trees often contain rules that prioritize flagging a potential vulnerability (e.g., a dictionary word or a pattern) over rewarding positive attributes like length. This "fail-secure" tendency is a desirable trait for a password strength scoring tool.

In summary, the 'medium' category is not merely an intermediate point but represents a complex classification zone where the model weighs competing evidence. The analysis shows that our Random Forest model effectively uses the hybrid feature set to identify passwords that appear strong to traditional checkers but harbor the nuanced vulnerabilities that modern attackers actively exploit.

### 4.5. Feature Importance Using SelectKBest

To pinpoint the most crucial features for classifying password strength, we employed SelectKBest with the ANOVA F-value scoring function, which assesses how effectively each feature differentiates between strength classes. Our results revealed that character variety scores and entropy were the most influential predictors, highlighting their essential role in distinguishing weak from strong passwords due to their inherent resistance to brute-force and dictionary attacks. Pattern detection emerged as the fifth most significant feature, emphasizing that even seemingly complex passwords can be vulnerable if they contain predictable structural patterns like "123456" or "password123". While pattern detection was a significant feature, its lower ranking relative to entropy and character variety suggests that many predictable patterns (e.g., '123456') are also inherently captured by having low entropy and low character variety. In contrast, basic metrics received the lowest scores, indicating that simple character counts alone are insufficient for accurate strength assessment, a finding consistent with criticisms of traditional LUDS (Lowercase, Uppercase, Digits, Symbols) password policies. This analysis underscores the importance of a hybrid feature engineering approach, combining classical metrics like entropy with semantic and structural checks such as pattern detection, to identify nuanced vulnerabilities that rule-based systems often miss.

### 5. CONCLUSION AND FUTURE WORK

While our model demonstrates high accuracy, we acknowledge several limitations in this study. First, our model's performance is contingent on the quality of the ground-truth labels from the 000WebHostLeak dataset, for which the original labeling methodology is unknown. Second, our semantic analysis relied on a limited blacklist of 197 common passwords, which may not detect less common but still vulnerable dictionary-based passwords. Finally, the model was trained on a single, albeit large, dataset, and its generalizability across different user populations and password creation contexts remains to be tested.

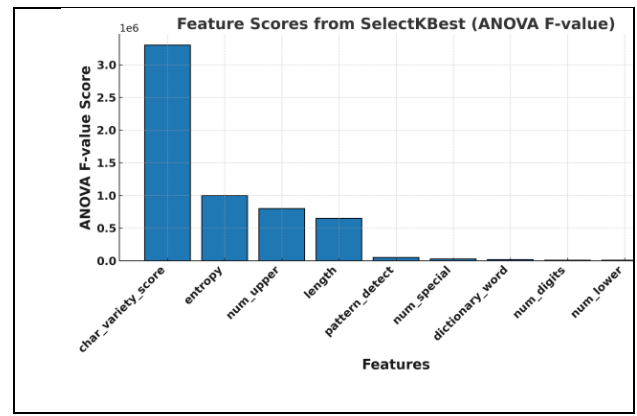

FIG. 4. FEATURES SCORES FROM SELECTKBEST

This paper has presented a robust methodology for password strength scoring. We applied machine learning techniques to develop a password strength scoring system that outperforms previous rule-based systems, which have grown increasingly ineffective in the face of growing cyber threats and user habits. Among the models tested, the Random Forest model demonstrated the highest accuracy and robustness, with a test accuracy of 99.12% and a weighted F1-score of 0.991, demonstrating its high effectiveness. This strong performance results from our comprehensive and carefully engineered hybrid feature set, which captures nuanced vulnerabilities using measures like entropy, character variety, and pattern detection.

Our findings underscore the need for advanced analytical techniques to enhance cybersecurity. The proposed system provides a more adaptive and accurate method of determining password strength, directly addressing the shortcomings of static evaluation criteria. Although the study has a strong foundation, we believe further hyperparameter tuning and tighter control of data preprocessing could enhance its methodological rigor. Additionally, focusing on user education and testing the system's resilience to malicious attacks would greatly improve its real-world applicability and contribute significantly to continuing efforts to empower users to make informed security decisions. This study is a significant step towards developing high-performance password scorers appropriate for real-world implementation, eventually contributing to a reduction in password-based vulnerabilities and a more secure environment.

Building on these results, future work will proceed in several key directions. First, we plan to enhance the semantic analysis component by integrating comprehensive password dictionaries, such as the RockYou corpus, to improve the detection of dictionary-based threats. Second, we will evaluate the model's robustness and generalizability by testing it against other large-scale password datasets from different data breaches. Third, we intend to explore more advanced deep learning architectures, such as Transformers, which may better capture long-range dependencies within passwords. Fourth, a user study will be designed to assess the real-world usability and effectiveness of our scoring system in guiding users toward creating stronger and more secure passwords.

Finally, and critically, we will investigate the model's potential biases to ensure fairness and equitable security. This includes a thorough analysis of its performance on passwords from diverse linguistic origins, particularly those containing non-ASCII characters. Such an investigation is essential to validate the model's generalizability and ensure it provides reliable protection for a global user base, rather than being unintentionally biased towards English-language or ASCII-based password creation patterns.